\documentclass[doublecol]{epl2}

\usepackage{graphicx}
\usepackage{dcolumn}
\usepackage{bm}
\usepackage{subfigure}

\graphicspath{{figures/}}

\title{Some scale-free networks could be robust under the selective node attacks}
\shorttitle{Some scale-free networks could be robust under the selective node attacks}

\author{Bojin Zheng\inst{1,2} \and Dan Huang\inst{1} \and Deyi Li*\inst{3}\and  Guisheng Chen\inst{2,3} \and Wenfei Lan\inst{1}
}

\institute{
  \inst{1} College of Computer Science,
South-Central University for Nationalities, Wuhan 430074, China \\
  \inst{2} State key Lab. of  Networking and Switching Technology, Beijing University of Posts and Telecommunications, Beijing 100876, China\\
  \inst{3} School of Software, Tsinghua University, Beijing 100084, China
}


\abstract{
It is a mainstream idea that scale-free network would be fragile under the selective attacks. Internet is a typical scale-free network in the real world, but it never collapses under the selective attacks of computer viruses and hackers. This phenomenon is different from the deduction of the idea above because this idea assumes the same cost to delete an arbitrary node. Hence this paper discusses the behaviors of the scale-free network under the selective node attack with different cost. Through the experiments on five complex networks, we show that the scale-free network is possibly robust under the selective node attacks; furthermore, the more compact the network is, and the larger the average degree is, then the more robust the network is; With the same average degrees, the more compact the network is, the more robust the network is. This result would enrich the theory of the invulnerability of the network, and can be used to build the robust social, technological and biological networks, and also has the potential to find the target of drugs.
}

\begin{document}
\maketitle

The invulnerability of complex network is an important issue for various of real-world networks\cite{a28,a18,a26}, such as the social relationship network\cite{a5}, Internet\cite{a6}, the World Wide Web\cite{1}, and the neural networks\cite{a9} and so on. The knowledge on this topic may lead to build more robust social, biological and technological networks, to find the targets of drugs or to destroy the terrorist groups etc.. Moreover, the knowledge may lead to deeper understandings of the stability of complex systems.

The pioneer scientists Re\'{k}a Albert, Hawoong Jeong and Albert-L\'{a}szl\'{o} Barab\'{a}si proposed that the complex networks with scale-free property would be robust with random errors, but fragile under the selective attacks, that is, the robustness and the fragility coexist\cite{a28}. This conclusion can be explained as follows, since the scale-free networks have a mass of nodes with less connectivity and a small quantity of nodes with huge connectivity, the nodes with less connectivity would be chosen with a high probability when the nodes are randomly chosen to remove, such that the network would lose less edges and show large invulnerability; on the other hand, if the nodes are preferentially chosen, the nodes with large connectivity would be firstly removed, such that the network would be very fragile to collapse.

The following researches proved the idea of Albert et al.,
for examples, the studies on the protein network\cite{a12}, on WWW\cite{a13},on
the food chain network\cite{a14}, on the Email network\cite{a15},
on Internet\cite{a16} and so on. These researchers made the idea
as a common knowledge of the scientists in this field. Holme et al. \cite{a18} summarized that the selective attacks can be categorized as node attacks
and edge attacks, and for each class, there are four strategies, that is,
ID(Initial-Degree), RD(Recalculated-Degree), IB(Initial-Betweenness), RB(Recalculated-Betweenness), meaning that based on the initial graphs or current graphs, order the degree or the betweenness of the nodes/edges descendingly, and chose the best node one by one to remove. Holme et al. pointed out that scale-free networks are fragile under all the attack strategies.

However, some scale-free networks provide exceptional excuses.
For example, Internet is regarded as a typical scale-free network, but it has
surprising stability and rarely collapses. There are at least two reasons to support its strong robustness: first, this network survives under the selective attacks of the hackers. No hacker has achieved the boastful victory although they have tried some very important nodes; second, if one treats the attacks of the hackers as random attack, although we do not agree on this treatment, the spread of computer viruses would provide strong persuasion. For the nodes with large degree, they will be infected with large probability because of their numerous neighborhoods. From the view of statistics, this is exactly the selective attacks. Another example is the military networks. Due to the hierarchy of military groups, the scale-free networks can be constructed\cite{a27,a29}. According to Albert et al.'s idea, the military networks would be very fragile and the strategy ``to catch bandits first catch the ringleader'' would be the best. But, if this idea holds, according to the evolutionary theory, the military groups would evolve into other forms to confront the selective/military attacks; on the other hand, the strategics would not be so colorful since only one strategy is enough.

Albert et al.'s idea assumed that the probability to remove any node or edge would be the same, or say, the cost is the same. Surprisingly, few works explored the world out of this hypothesis. Actually, no matter in Internet, or in military groups, the cost to remove a node would be certainly different. Moreover, the works mainly employ the degree attacks and the betweenness attacks. Considering that ``the important nodes should be important to the whole network'', the attack strategies would be based upon the definition of ``the importance of nodes''. Therefore, the closeness\cite{a19} or the other definition such as the definition in PageRank\cite{a21,23} and HITS\cite{a23} also can be used as an attack factor.

Thus, when the removal cost is different for the nodes with different importance, what will happen? We can image that a mass of man power and materials will be used to enhance the protection of the important nodes such that the important nodes could resist the attacks, thereby lead to the increase of the robustness of the whole network. Therefore, when considering the cost, the scale-free network may be robust. In this Letter, we try to explore the invulnerability of complex network with the consideration of the different costs to delete the nodes and find that some complex networks could be very robust under the selective attacks.

For the quantitative validation of the robustness of scale-free network under selective node attacks with the consideration of the cost, we define the removal cost (attack cost), the attack strategies, the attack effect based on the previous related works, and we find that the compactness and the average degree of scale-free network are related to the robustness, furthermore, the more compact the scale-free network is, and the larger the average degree is, then the more robust the network is; for the same average degree, the more compact the scale-free network is, the more robust the network is. Five scale-free networks are used to validate our idea.

%

Here we firstly define the relevant concepts. Since we only consider the selective node attacks, we define the attack as the removal of a set of nodes. Since the degree measure is the most natural measure to the property of nodes, we use a function of the degree of the set of nodes as the measure of the cost of an attack. That is, for given graph $G=<V,Edge>$ and the node set for the removal of nodes $Z$,

\begin{equation}\label{eq.cost}
  Cost(Z) = \sum\limits_{v \in Z}{f(Degree(v))} 
\end{equation}

Here, $f(x)$ can be defined as various functions, for example, $f(x) = x^2$. Under this circumstance, since the attack cost of the nodes with higher degree would be remarkably large, the best strategy would be to remove the nodes with smaller degrees firstly. If $f(x) = 1/x$, since the attack cost of important nodes would be smaller, it would be efficient to remove the nodes with large degree. If $f(x)= x^{0}$, that is, $f(x)= 1$, the results would be the same as Barab\'{a}si's. The fairest formation would be $f(x) = x$, here the cost is equivalent to the degree, it also is the simplest formation. This Letter suggests this function.

For any given network, there exists an upper limitation when all the edges are removed. We normalize the cost and have(denoted the normalized cost as $C(Z)$),

\begin{equation}\label{eq.normcost}
C(Z) = \frac{{Cost(Z)}}{{Cost(V)}}
\end{equation}

As to the attack strategies, due to the fact that the important nodes would be important to the whole network, we can design the attack strategies according to the definitions of the node importance. Considering the definitions in Social Network\cite{a19}, we focus on three strategies, IB (Initial-Betweenness), ID (Initial-Degree) and IC (Initial-Closeness). Focusing on the selective node attacks, the important nodes would be preferentially removed.

As to the attack effects, the researchers have proposed various measures. Commonly, it can be measured by the difference of the performance of the tested network before or after the attack, such as average geodesic length, average inverse geodesic length and the size of giant component\cite{a18}. We use the size of giant component after the attack $\tilde{S}(Z)$ as the measure indicator, and considering universal and convenient comparisons, we use the normalized size $E(Z)$, where,

\begin{equation}\label{eq.normeffect}
E(Z)=\frac{\tilde{S}(Z)}{\|V\|}
\end{equation}

Considering that there exist many kinds of attack strategies, a network is said to be robust unless
it can resist multiple kinds of attack strategies. If considering for the degree and the closeness only,
assume that the degree and the closeness is negatively correlated, then, the important nodes (with larger closeness) can be removed with smaller
cost (equivalent to the degrees), such that the network is fragile. Therefore, the compactness of the network, or, every node importance measure would
be positively related to the degree, is the necessary condition for the robustness
of the network.

Furthermore, considering that the networks with larger average degree would have high probability to prevent from the collapse, that is, they may not collapse after the removal of
one or multiple most important nodes, we will investigate the relationship between the robustness and the average degree.

\begin{figure*}[hptb]
\subfigure[CSF-Compact]{
\label{fig:subfig:a:t} 
\includegraphics[width=3.54cm,height=2.8cm]{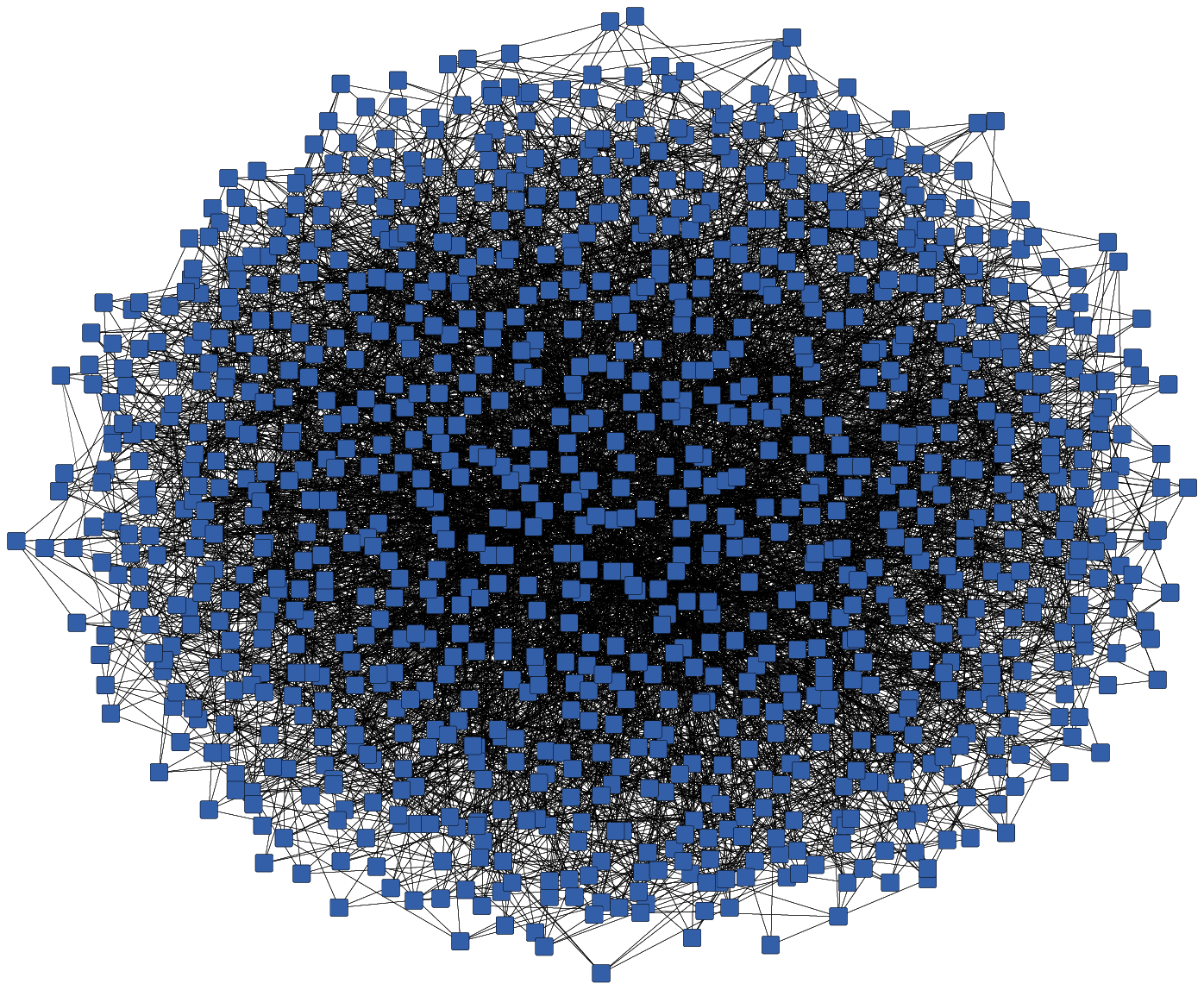}}
\hspace{-0.3cm}
\subfigure[CSF-NonCompact]{
\label{fig:subfig:b:t} 
\includegraphics[width=3.54cm,height=2.8cm]{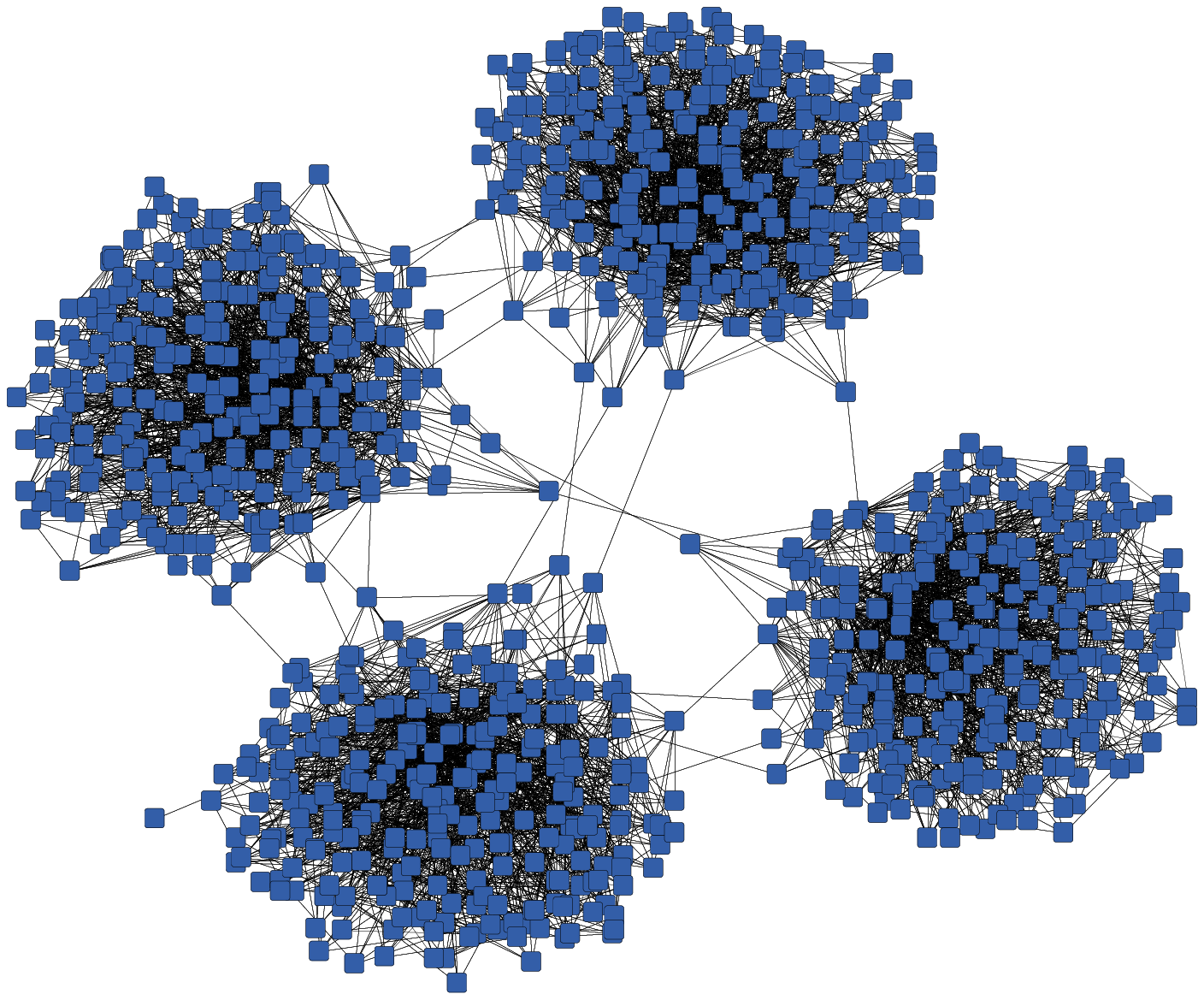}}
\hspace{-0.3cm}
\subfigure[Polbooks]{
\label{fig:subfig:c:t} 
\includegraphics[width=3.54cm,height=2.8cm]{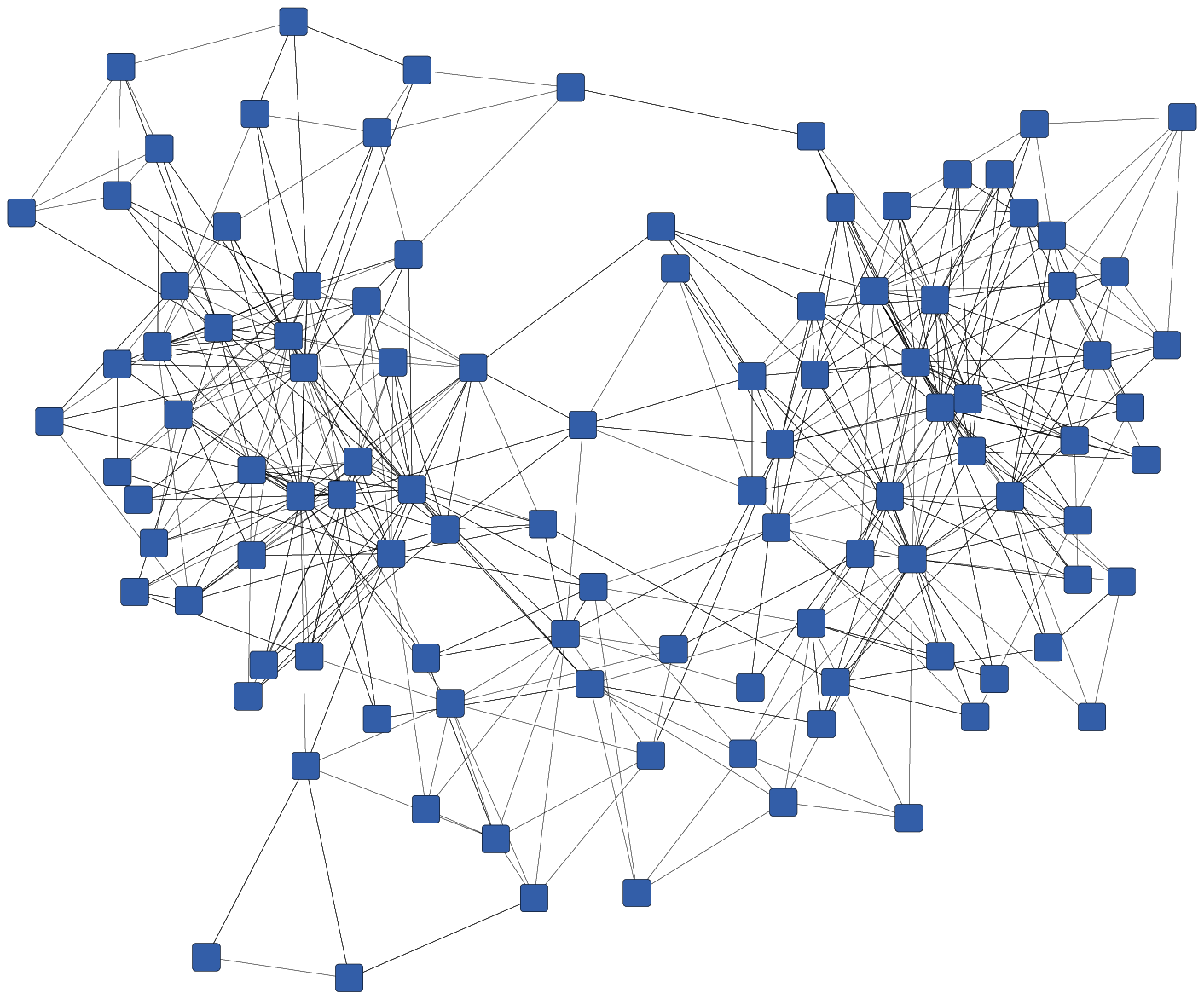}}
\hspace{-0.3cm}
\subfigure[Protein]{
\label{fig:subfig:d:t} 
\includegraphics[width=3.54cm,height=2.8cm]{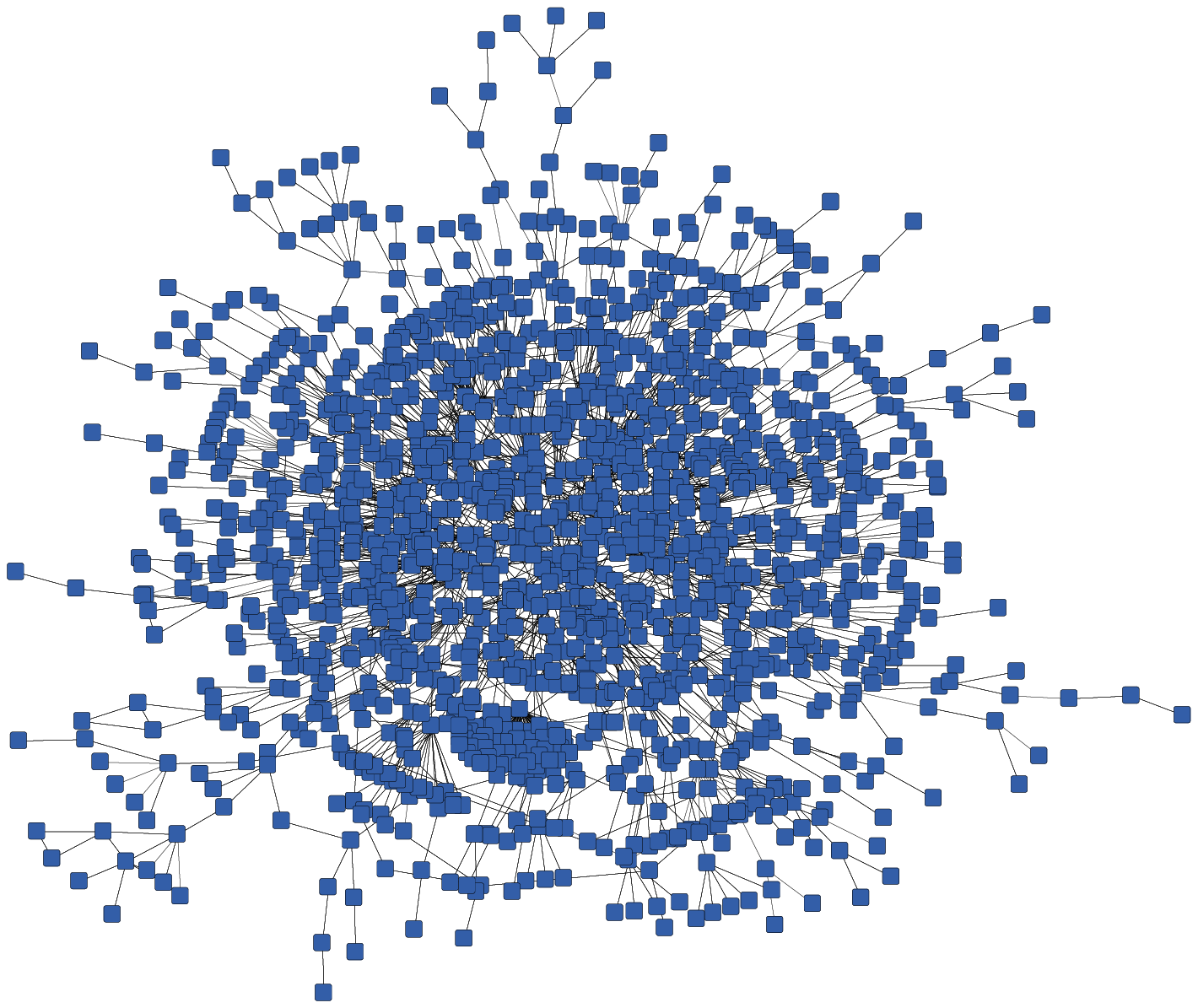}}
\hspace{-0.3cm}
\subfigure[Science]{
\label{fig:subfig:e:t} 
\includegraphics[width=3.54cm,height=2.8cm]{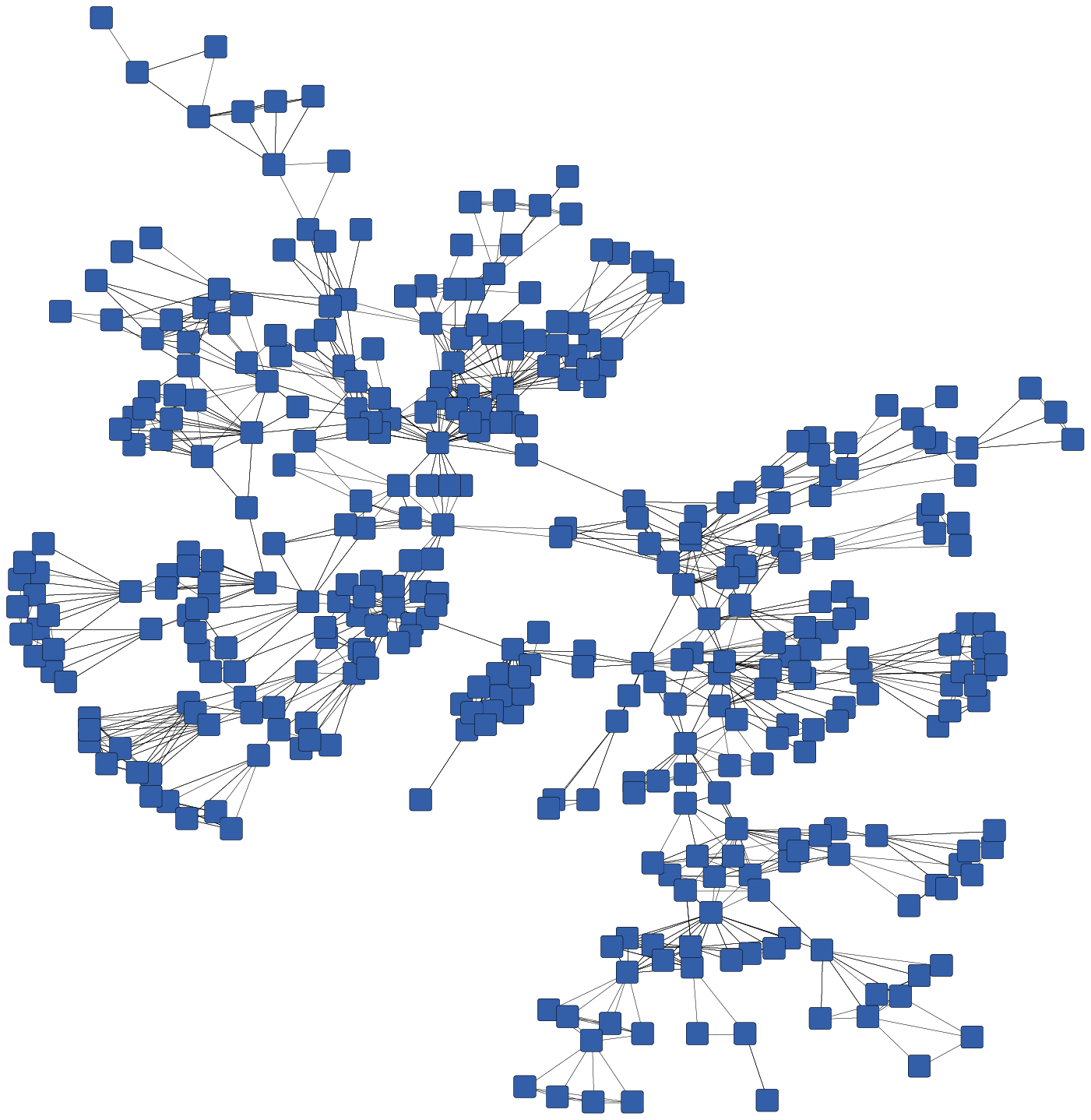}}
\caption{The Topographies of the Selected Networks}
\label{fig.netpictures} 
\end{figure*}

We use five networks to explore their behaviors under the selective node attacks, whose topographies are drawn in Fig.\ref{fig.netpictures}. Two of them are generated by CSF algorithm\cite{a18}, one is compact and shown as Fig.\ref{fig:subfig:a:t}, the other is non-compact and shown as Fig.\ref{fig:subfig:b:t}.
Three widely-used real-world networks, the protein network\cite{a13} shown as Fig.\ref{fig:subfig:c:t}
, the political books network (Polbooks\footnote{This network was edited by Valdis Krebs and can be downloaded from MEJ Newman's website.}) shown as Fig.\ref{fig:subfig:d:t} and the scientific collaboration network\cite{a2} shown as Fig.\ref{fig:subfig:e:t}, are chosen.

\begin{figure*}[hptb]
\subfigure[CSF-Compact]{
\label{fig:subfig:a:d} 
\includegraphics[width=3.54cm,height=2.8cm]{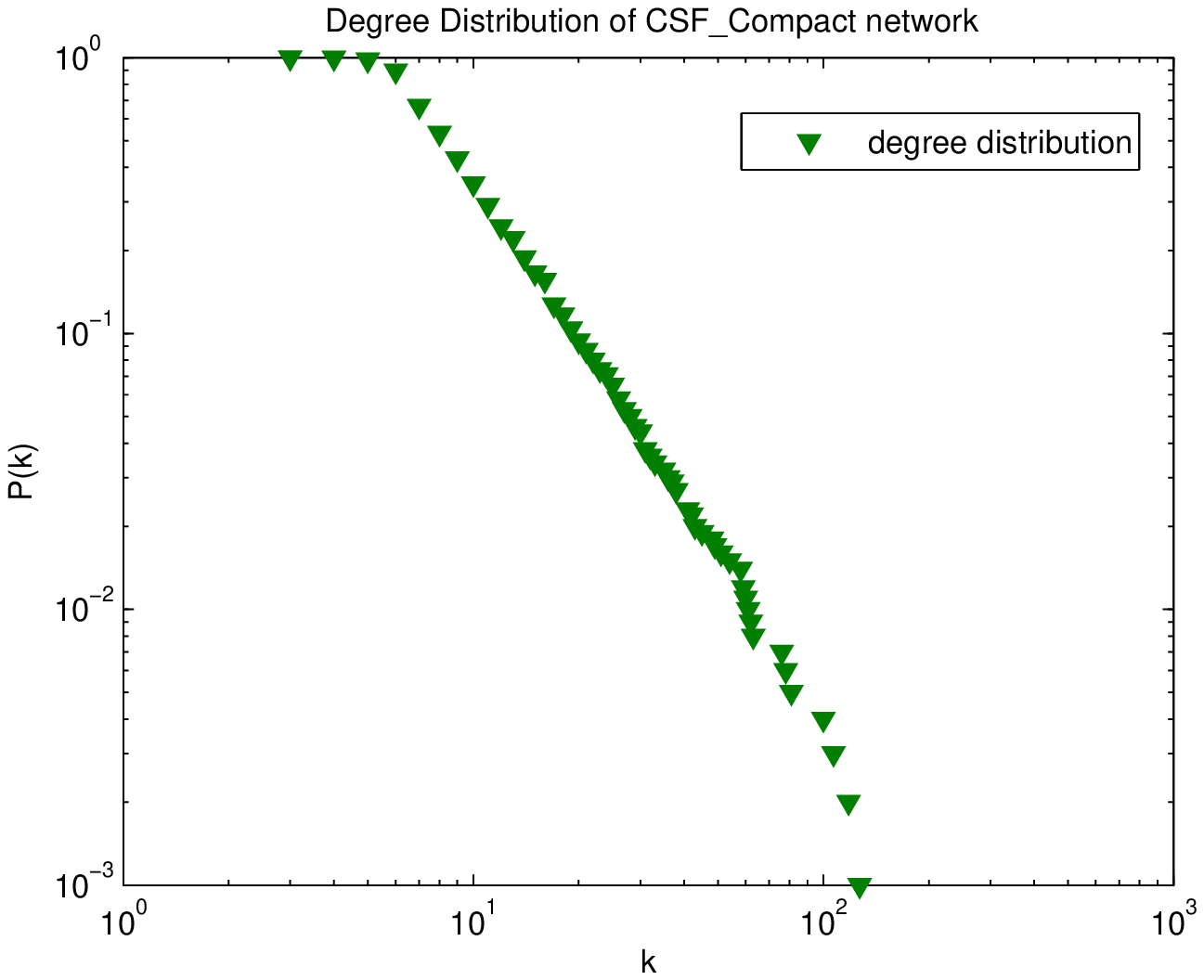}}
\hspace{-0.3cm}
\subfigure[CSF-NonCompact]{
\label{fig:subfig:b:d} 
\includegraphics[width=3.54cm,height=2.8cm]{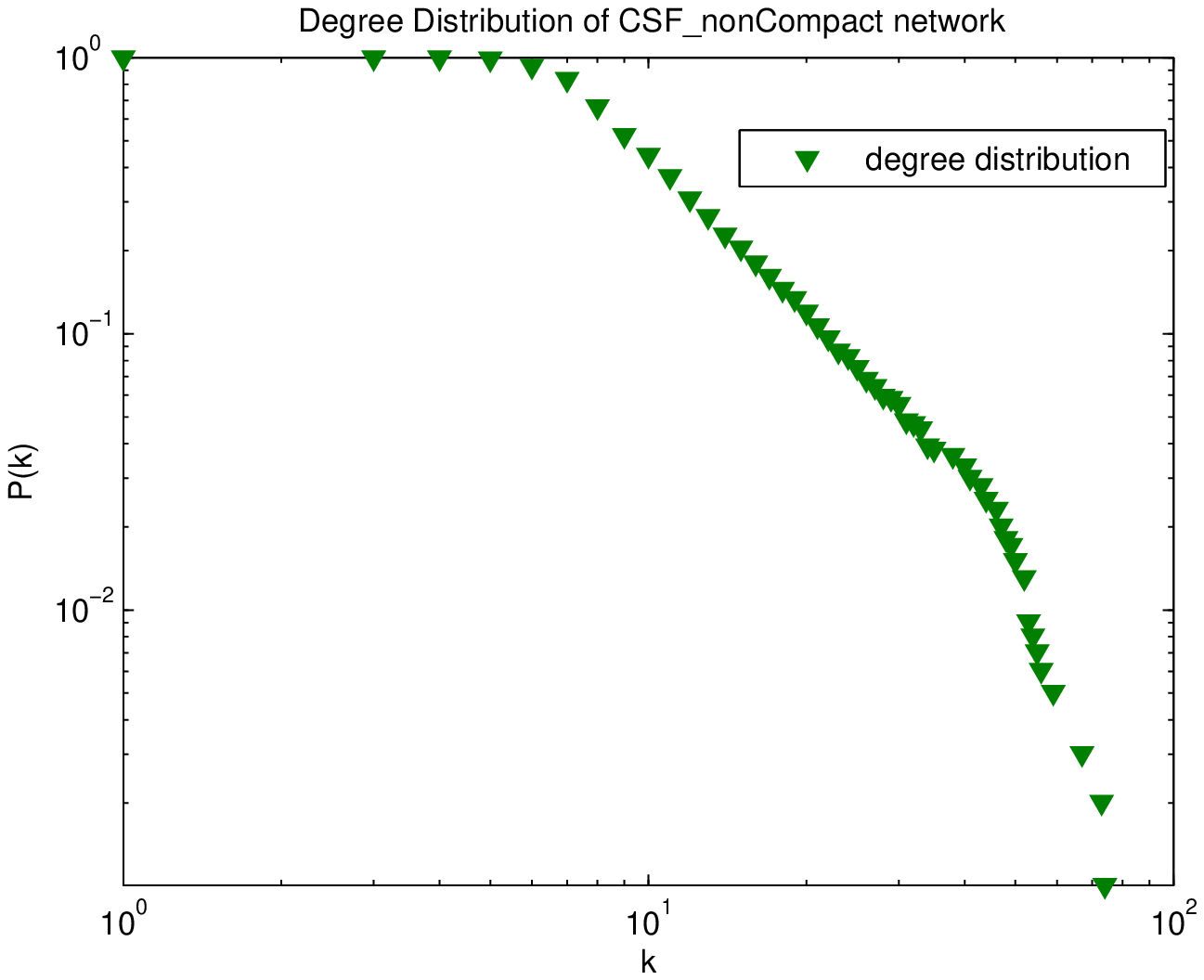}}
\hspace{-0.3cm}
\subfigure[Polbooks]{
\label{fig:subfig:c:d} 
\includegraphics[width=3.54cm,height=2.8cm]{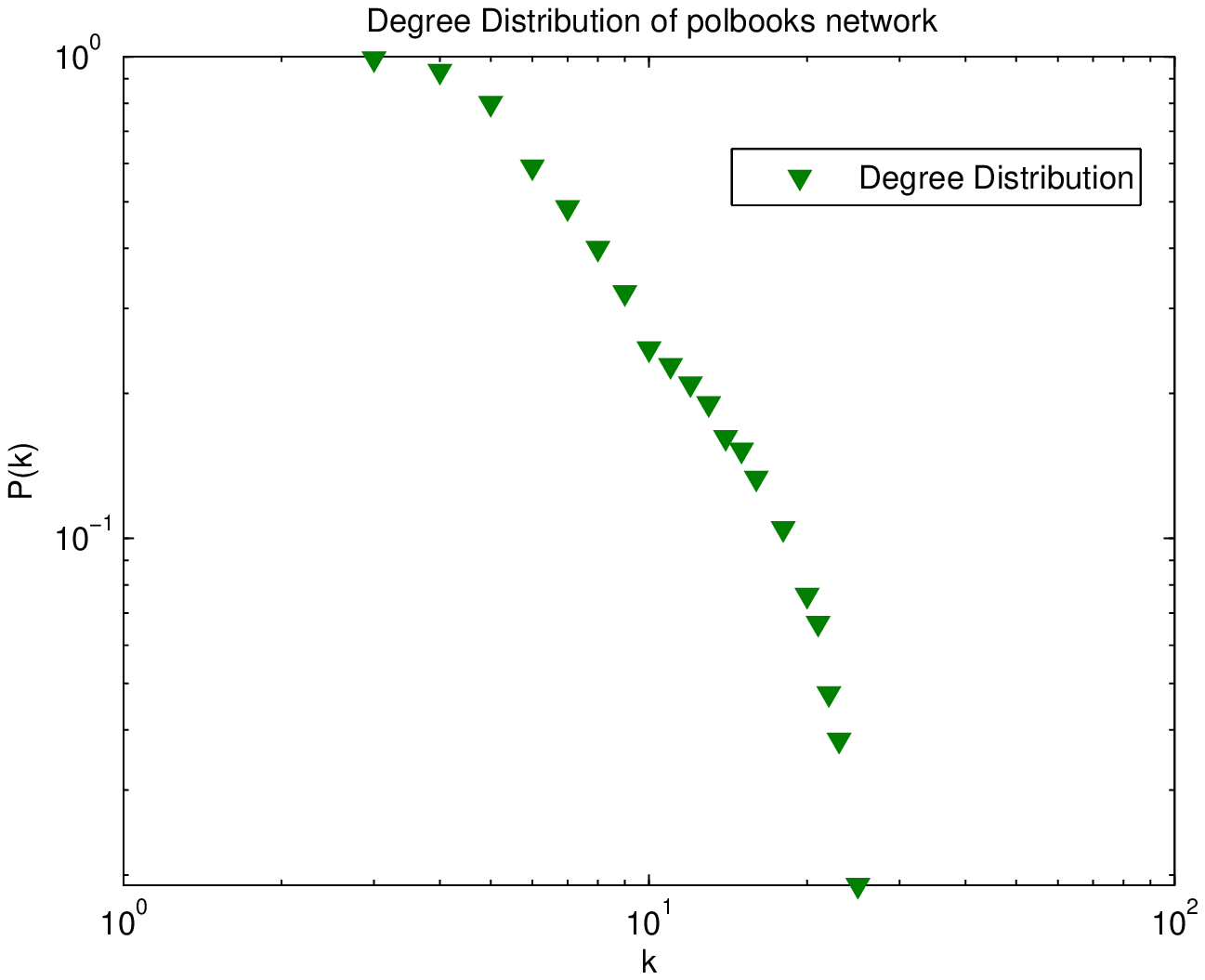}}
\hspace{-0.3cm}
\subfigure[Protein]{
\label{fig:subfig:d:d} 
\includegraphics[width=3.54cm,height=2.8cm]{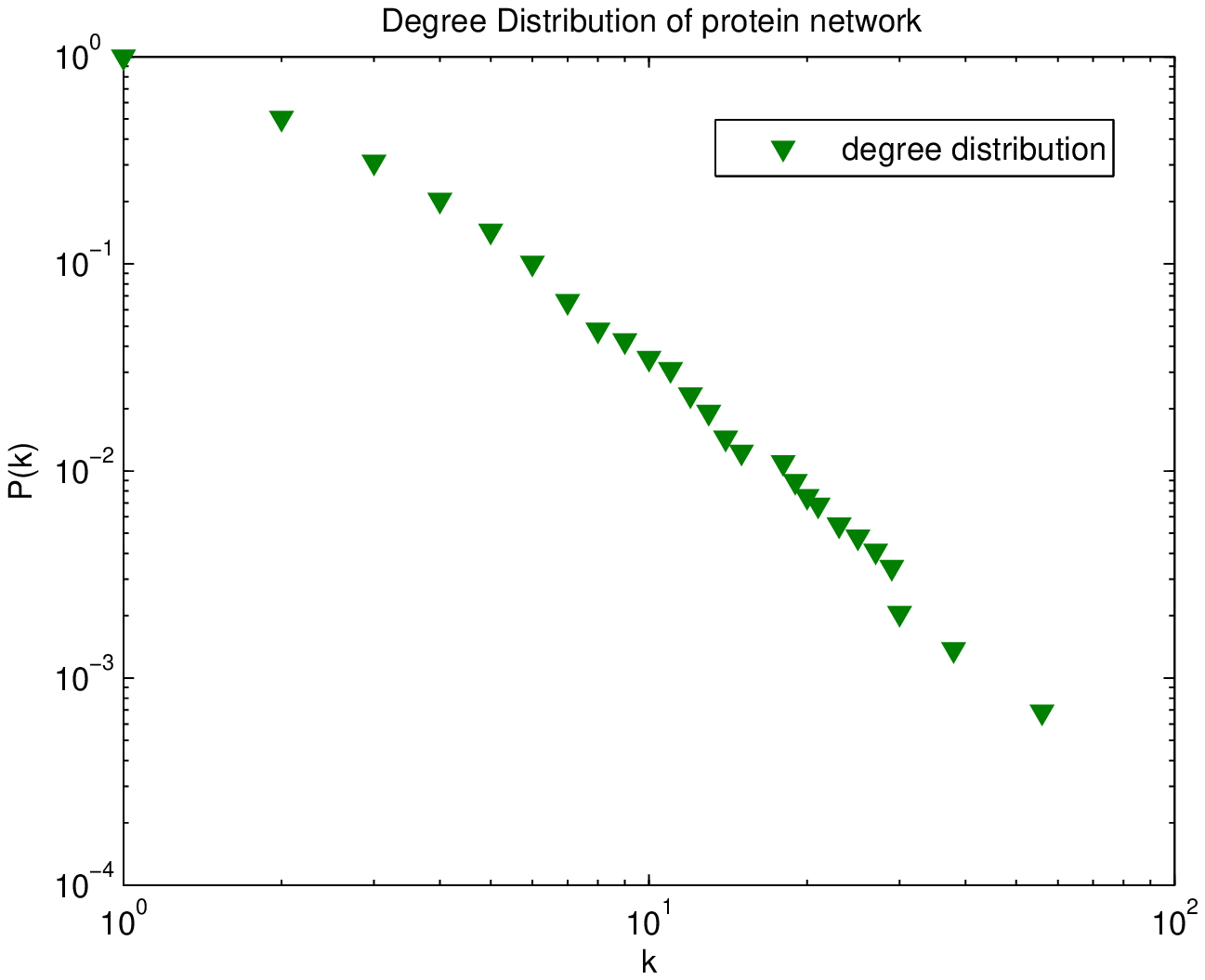}}
\hspace{-0.3cm}
\subfigure[Science]{
\label{fig:subfig:e:d} 
\includegraphics[width=3.54cm,height=2.8cm]{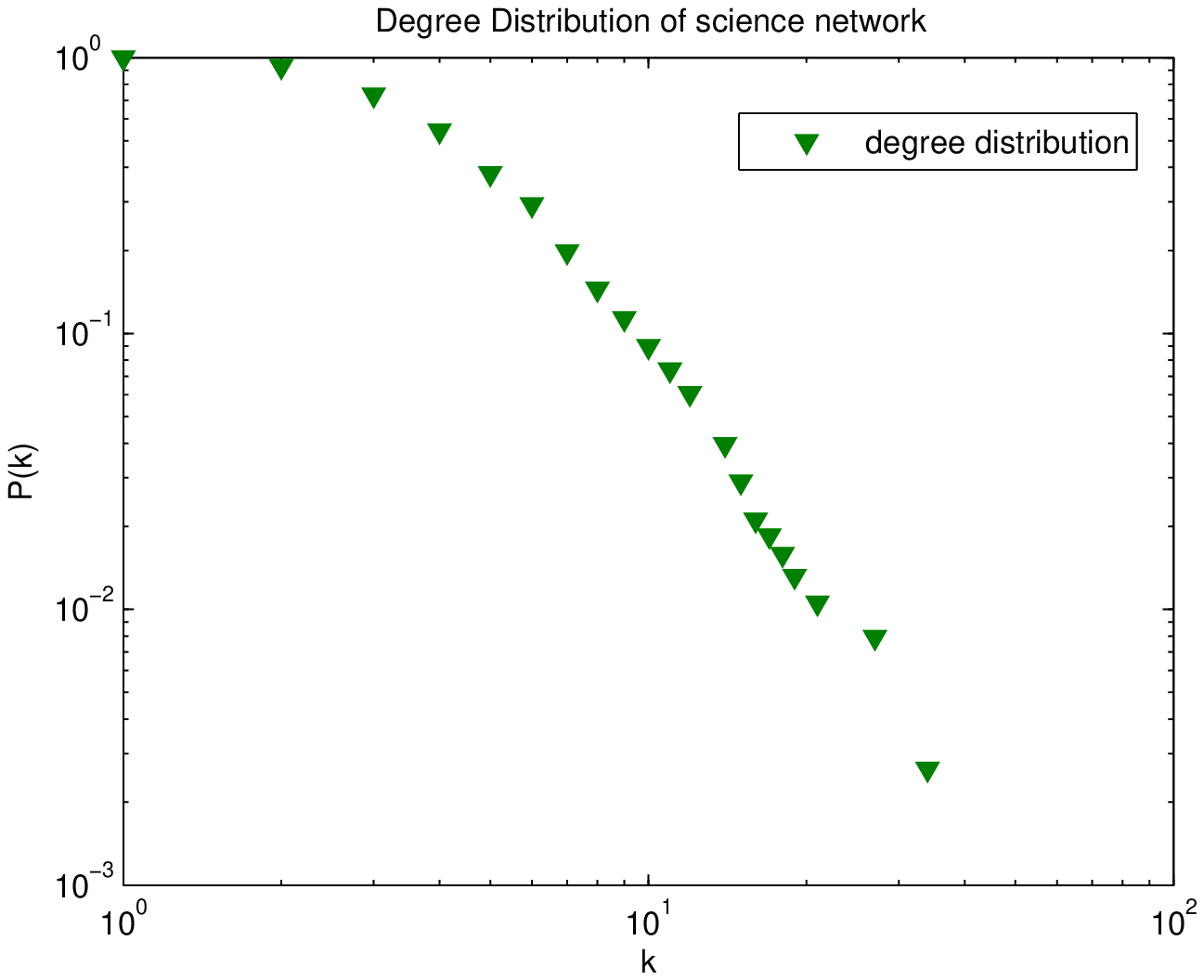}}
\caption{The Cumulative Degree Distributions of the Selected Networks. The X-axes are random variants on degree, and the Y-axes are the cumulative probability of random variants.}
\label{fig.DegreePictures} 
\end{figure*}

Since this Letter only focuses on the scale-free network, we need to
validate the scale-free property of the selected networks.  The common method is to check the networks if their degree distribution will satisfy $P(k) \sim k^{-\gamma}$, here $\gamma$ is a constant and k is a random variant on the degree, when plotting the equation in the log-log coordination, the curve looks a straight line.
We plot their degree distributions in Fig. \ref{fig.DegreePictures}. From this figure, we can see that the curves approximate the straight line in the log-log coordination. Considering that this method is qualitative, we also check their scale-free property by the programme written by Clauset et al.\cite{a32}.

To measure the correlated coefficients between the degree and the betweenness or the closeness, we use the Pearson's correlation coefficient, which belongs to $[-1,1]$ and $-1$ indicates the minimal negative correlation, and $1$ indicates the maximal positive correlation. The coefficients and some statistic values of the selected networks are listed in Tab. \ref{tab::Parameters}.

\begin{table}
\caption{\label{tab::Parameters}The statistic values of the selected networks. $N$ is the number of nodes, $Vx$ is the number of vertices, $\overline D $ is the average degree, $\delta_b$ is the Pearson's correlation coefficient between the degree and the betweenness, $\delta_c$ is the Pearson's correlation coefficient between the degree and the closeness.  }
\begin{tabular}
{cccccc}
\hline
Network& $N$ & $Vx$ &$\overline D $& $\delta_b$&$\delta_c$\\
\hline
CSF-Compact& 1000 & 5545&11.09&0.94&0.58 \\
Polbooks&105&441&8.40&0.70&0.58 \\
CSF-NonCompact&992&5948&12.0&0.45&0.62 \\
Protein&1458&1948& 2.67& 0.85& 0.42 \\
Science&379&914& 4.82& 0.69& 0.35 \\
\hline
\end{tabular}
\end{table}

\begin{figure*}[hptb]
\subfigure[CSF-Compact]{
\label{fig:subfig:a:e} 
\includegraphics[width=3.54cm,height=2.8cm]{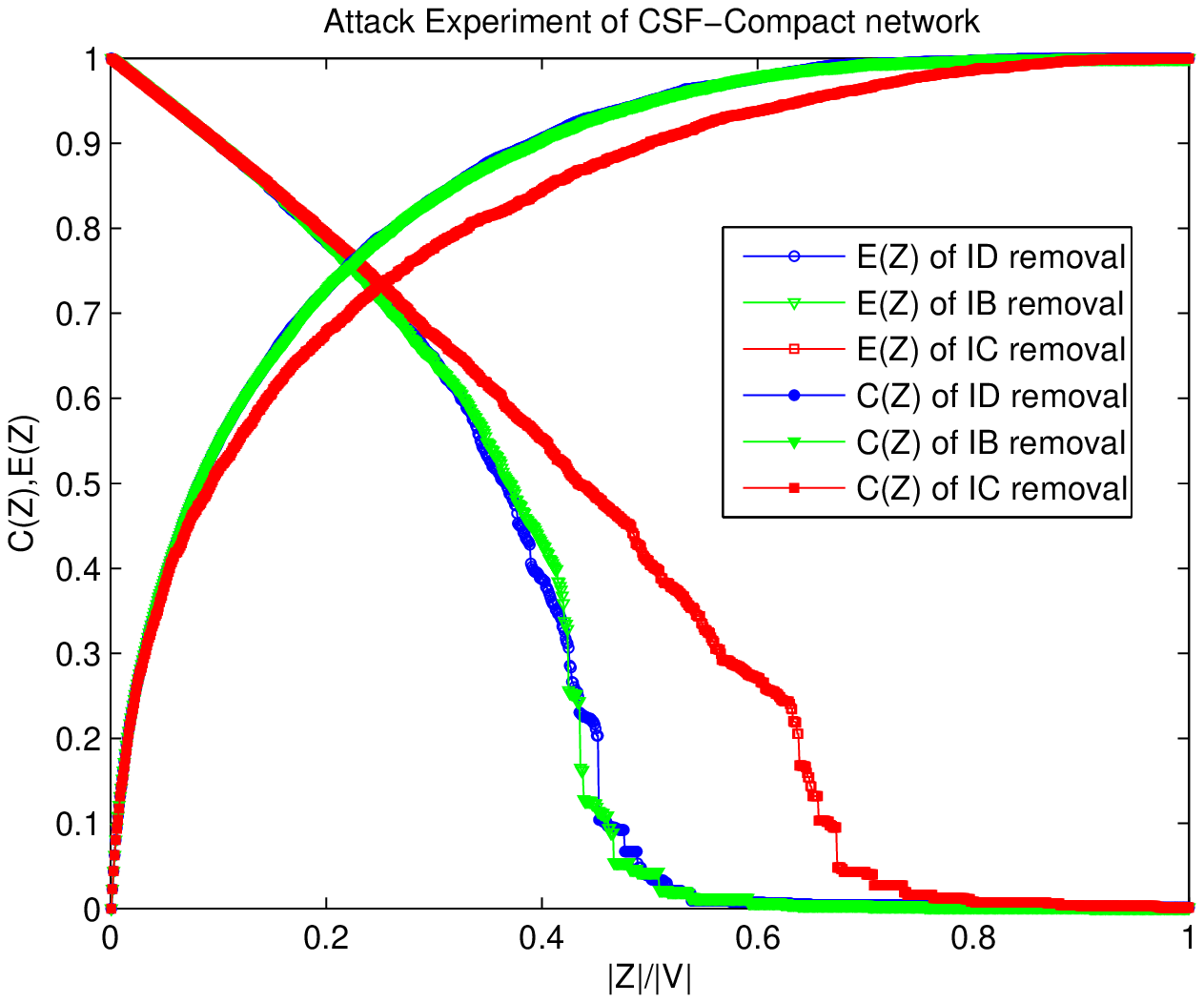}}
\hspace{-0.3cm}
\subfigure[CSF-NonCompact]{
\label{fig:subfig:b:e} 
\includegraphics[width=3.54cm,height=2.8cm]{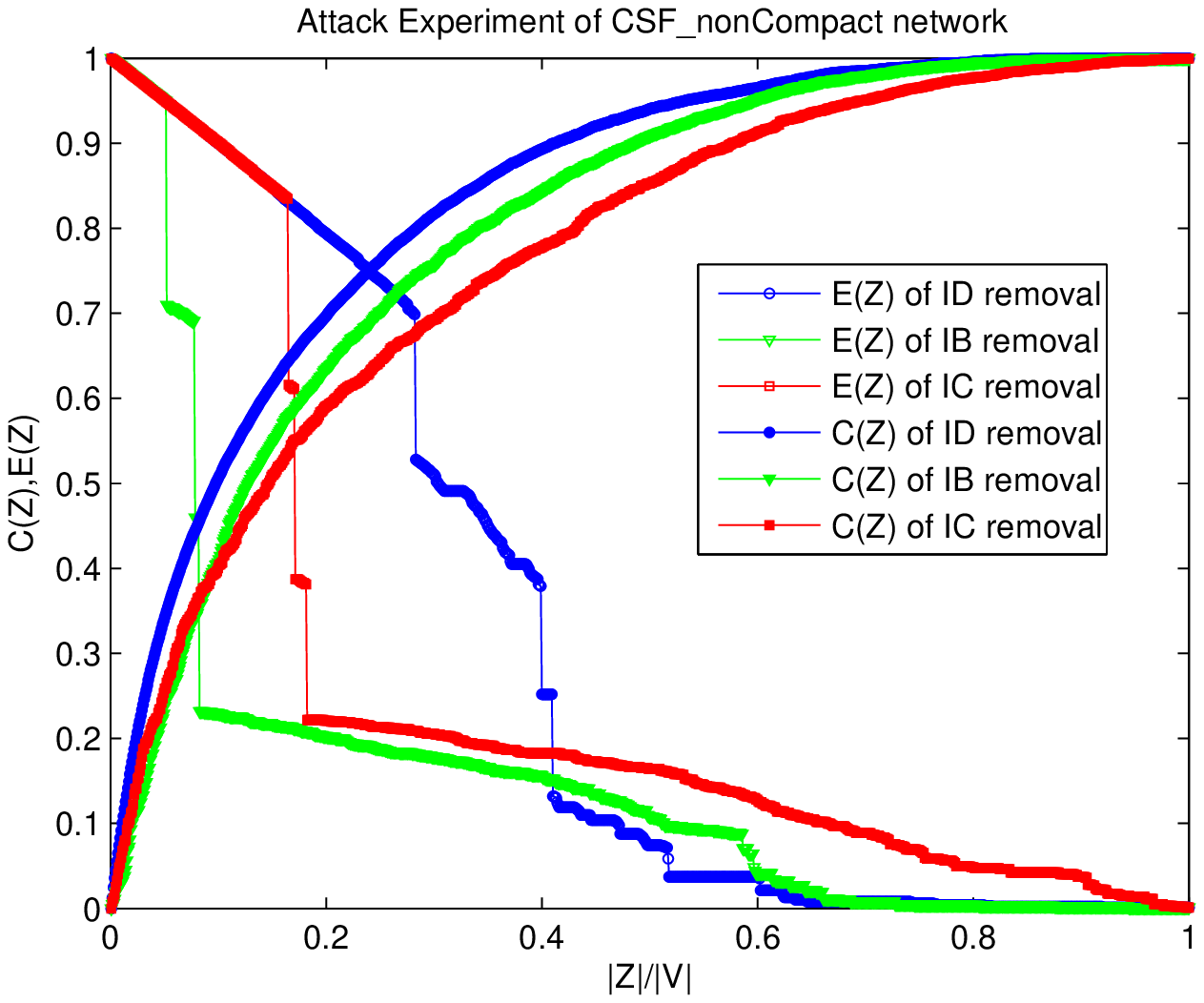}}
\hspace{-0.3cm}
\subfigure[Polbooks]{
\label{fig:subfig:c:e} 
\includegraphics[width=3.54cm,height=2.8cm]{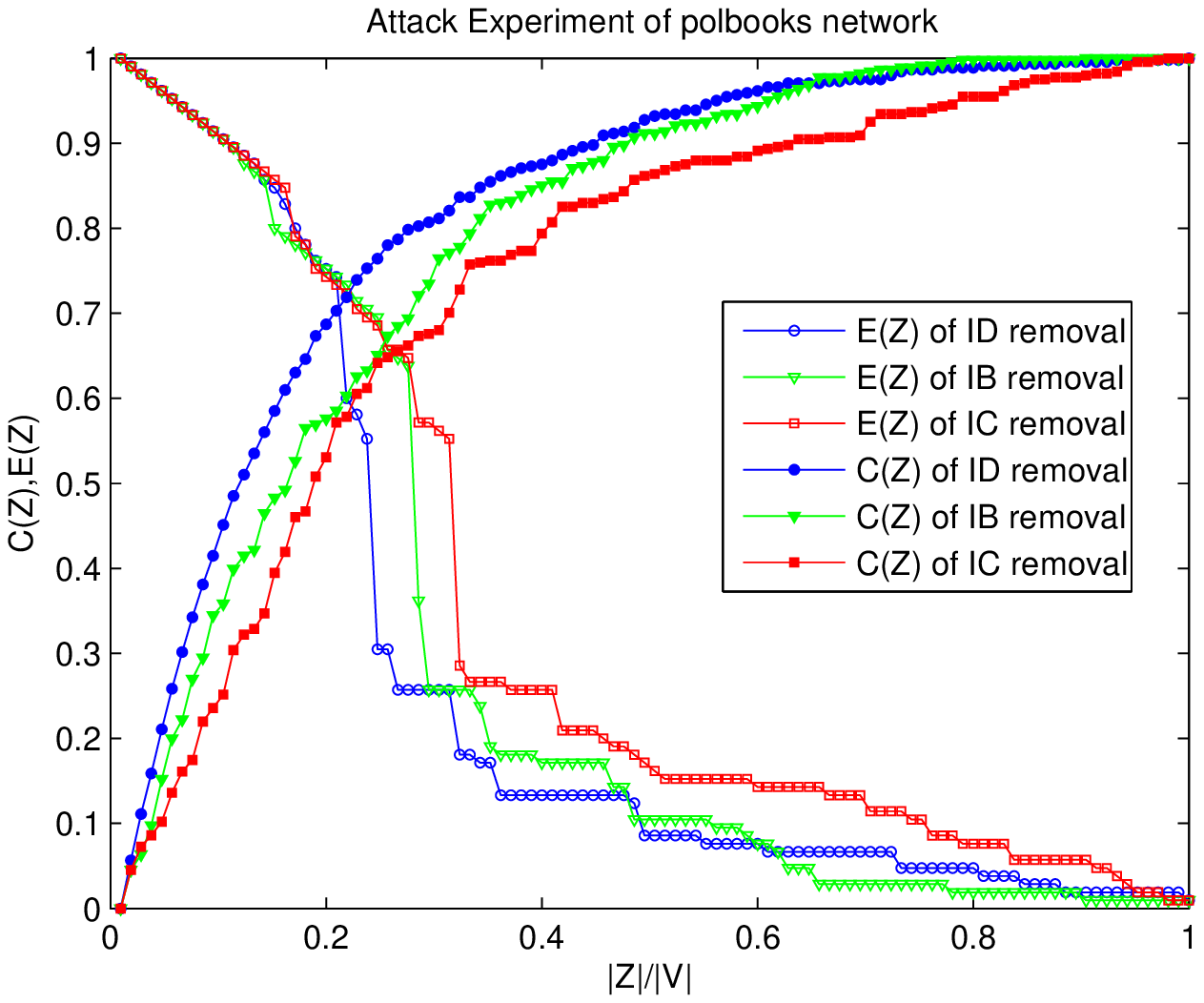}}
\hspace{-0.3cm}
\subfigure[Protein]{
\label{fig:subfig:d:e} 
\includegraphics[width=3.54cm,height=2.8cm]{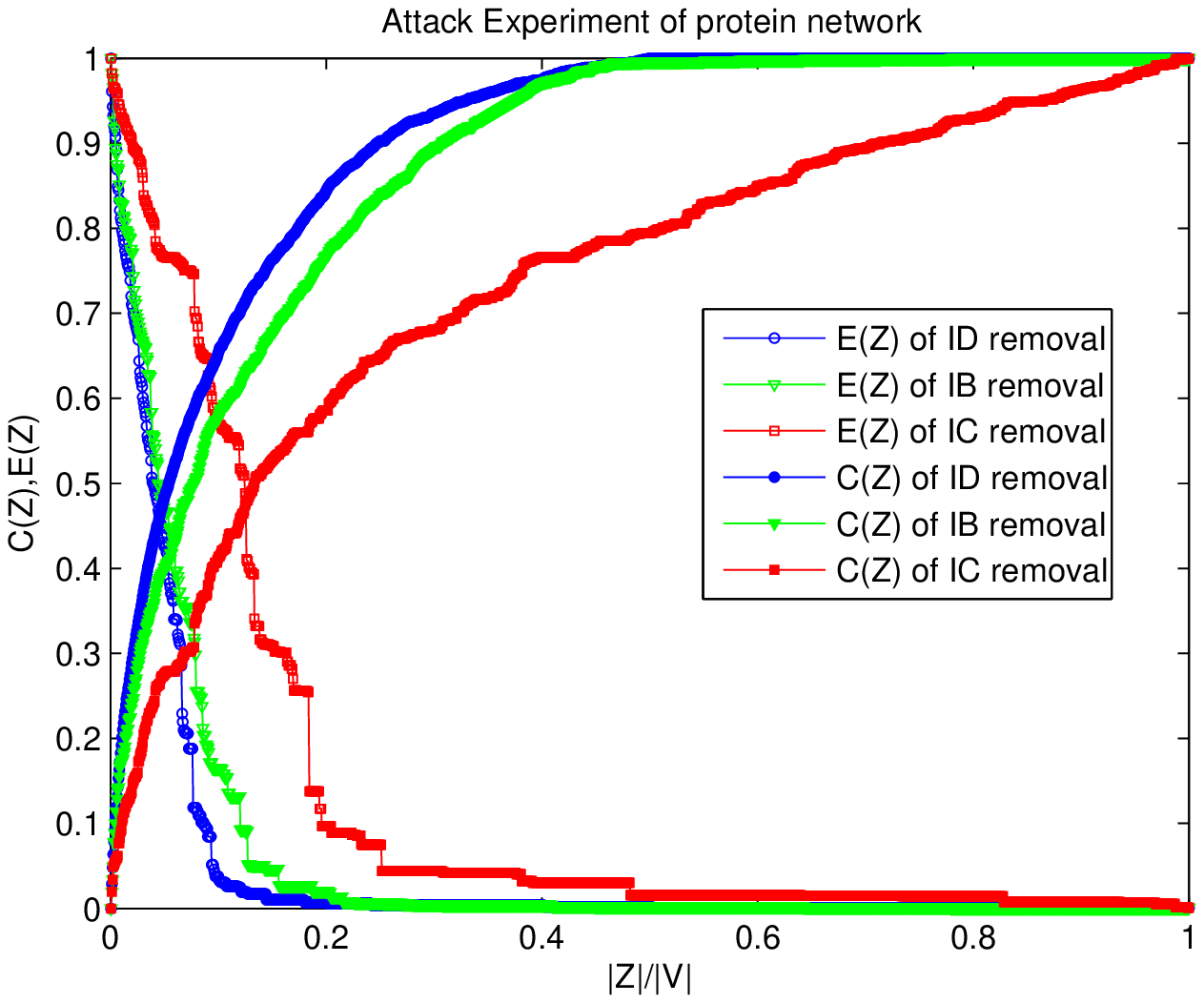}}
\hspace{-0.3cm}
\subfigure[Science]{
\label{fig:subfig:e:e} 
\includegraphics[width=3.54cm,height=2.8cm]{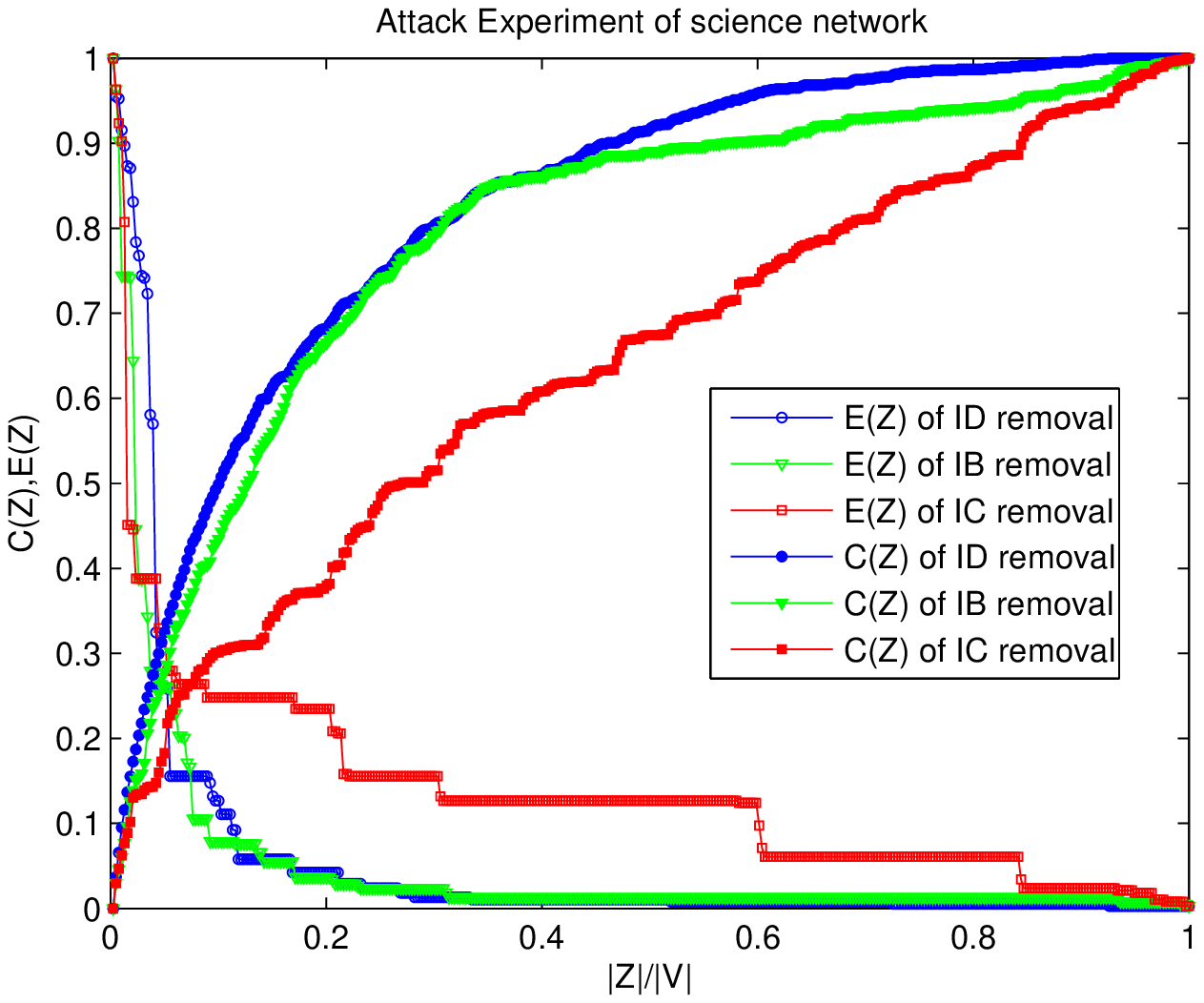}}
\caption{The Attack Experiments on the Selected Networks. The X-axes are the normalized size of the set responding to the attacks and the Y-axes are the responding attack cost $C(Z)$ and attack effect $E(Z)$. In every subfigures, better attack effect responds more cost.}
\label{fig.ExpPictures} 
\end{figure*}

We choose IB, ID, IC as the attack strategies to carry out the experiments. For any given attack, that is, the set $Z$ of nodes to be removed, we plot the corresponding normalized cost $C(Z)$ and the corresponding normalized attack effect $E(Z)$ under these strategies as Fig. \ref{fig.ExpPictures}. To avoid the distortion of the curves, the normalized size of the set $Z$ is used. Just the same as the previous works, $E(Z)$ drops down drastically for all these networks, meaning that the networks are very fragile. But we can also notice that $C(Z)$ increases drastically simultaneously. Therefore, when the cost is taken into consideration, we can not conclude that the networks are fragile.

\begin{figure*}[hptb]
\subfigure[CSF-Compact]{
\label{fig:subfig:a:r} 
\includegraphics[width=3.54cm,height=2.8cm]{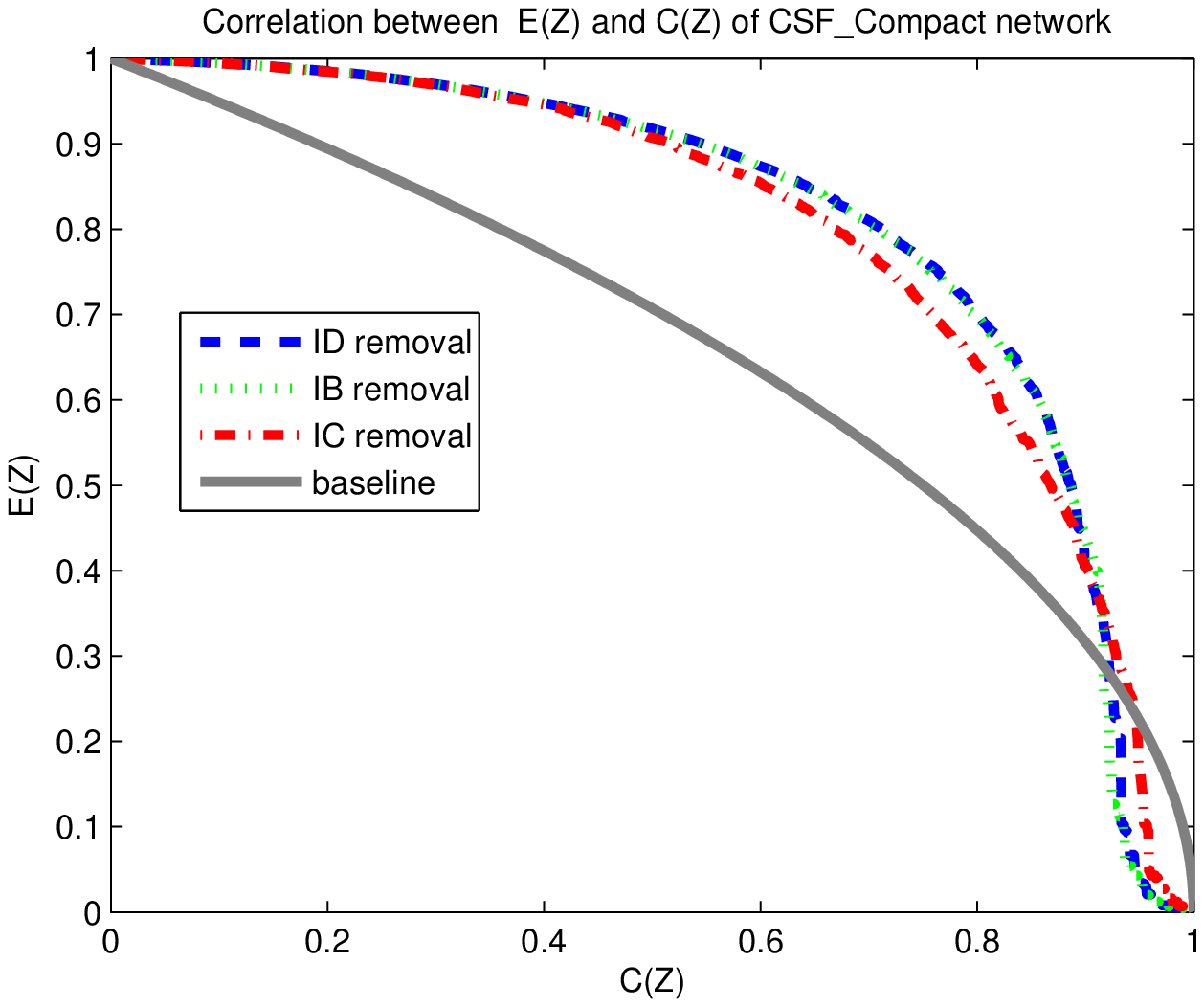}}
\hspace{-0.3cm}
\subfigure[CSF-NonCompact]{
\label{fig:subfig:b:r} 
\includegraphics[width=3.54cm,height=2.8cm]{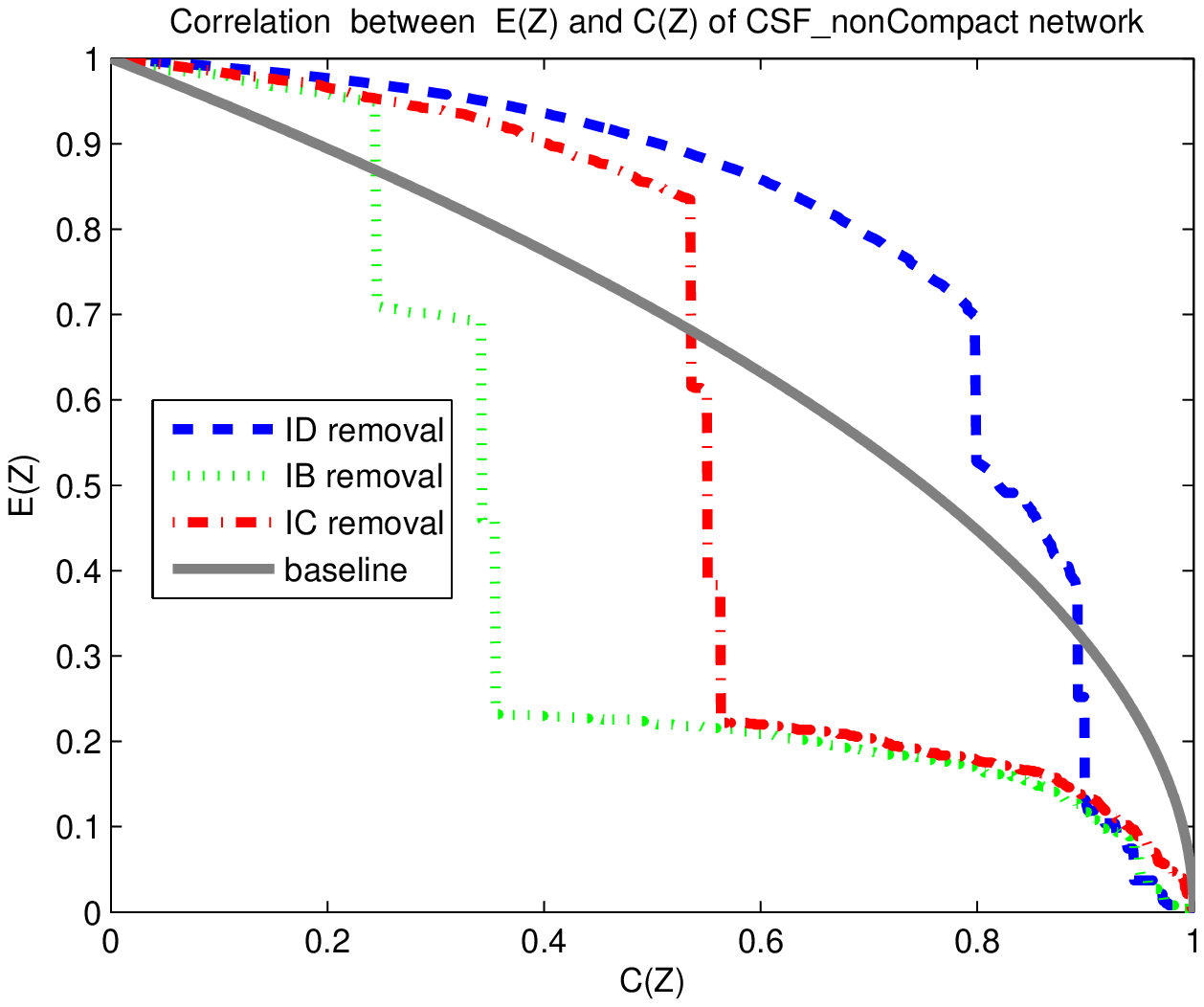}}
\hspace{-0.3cm}
\subfigure[Polbooks]{
\label{fig:subfig:c:r} 
\includegraphics[width=3.54cm,height=2.8cm]{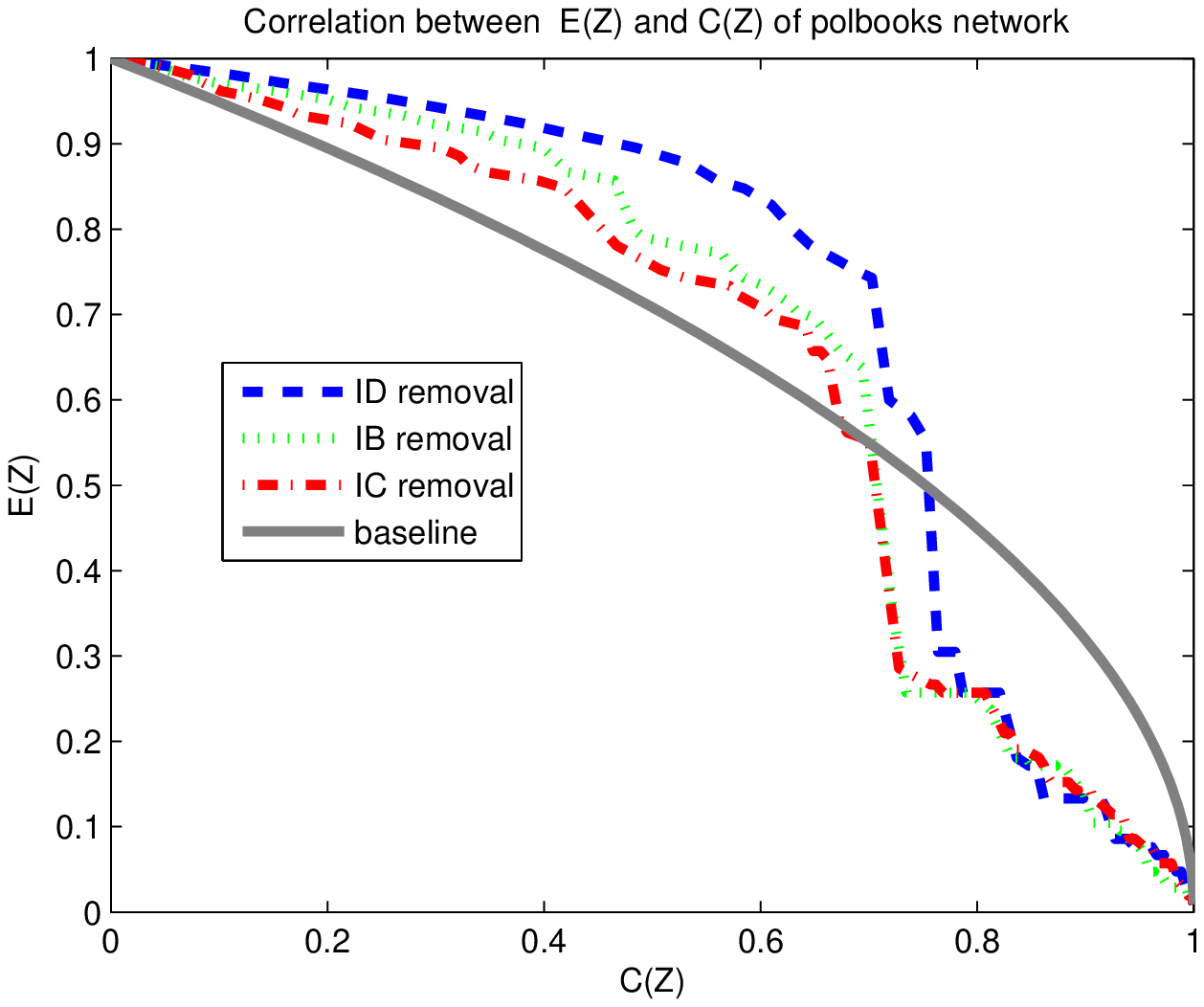}}
\hspace{-0.3cm}
\subfigure[Protein]{
\label{fig:subfig:d:r} 
\includegraphics[width=3.54cm,height=2.8cm]{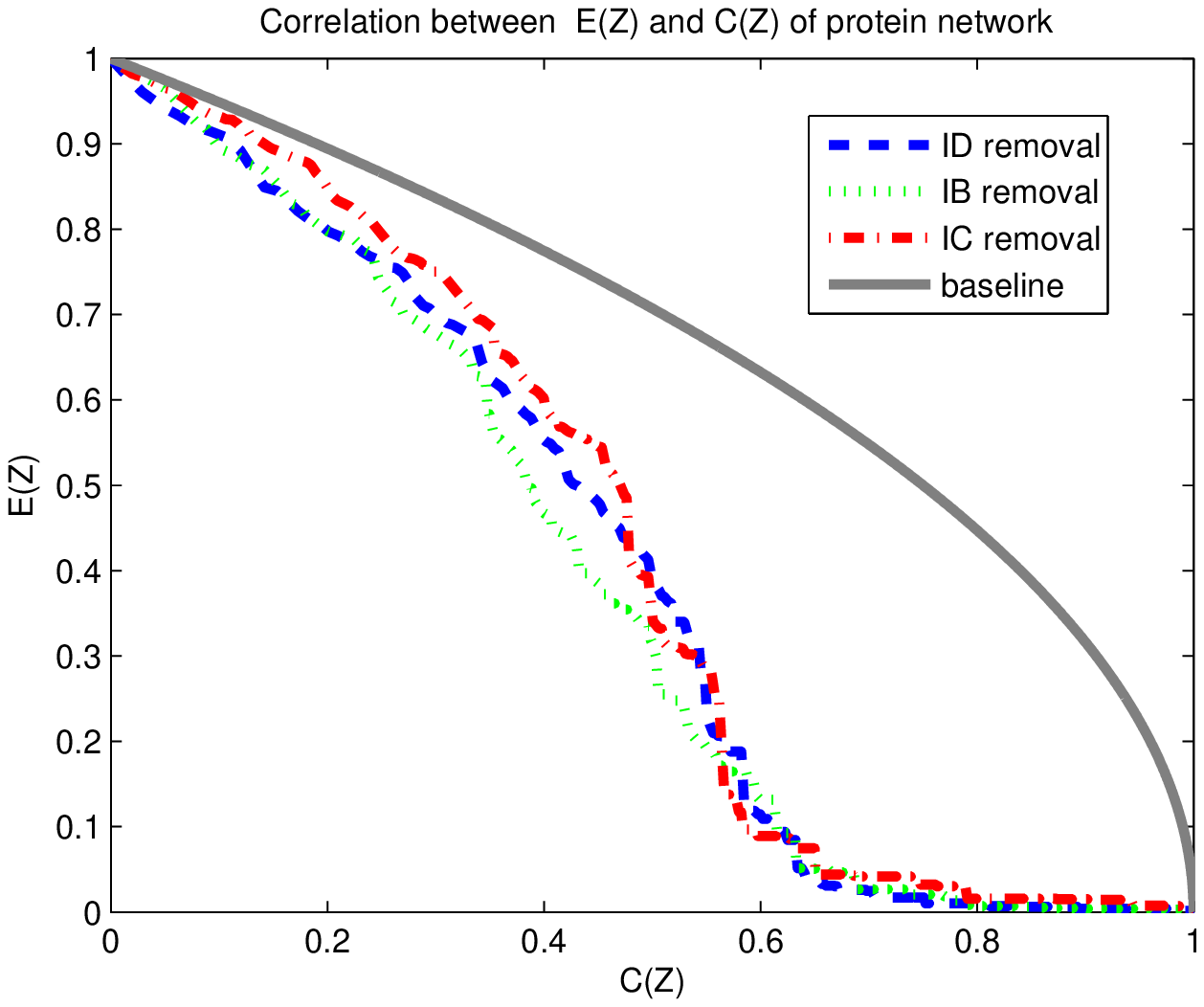}}
\hspace{-0.3cm}
\subfigure[Science]{
\label{fig:subfig:e:r} 
\includegraphics[width=3.54cm,height=2.8cm]{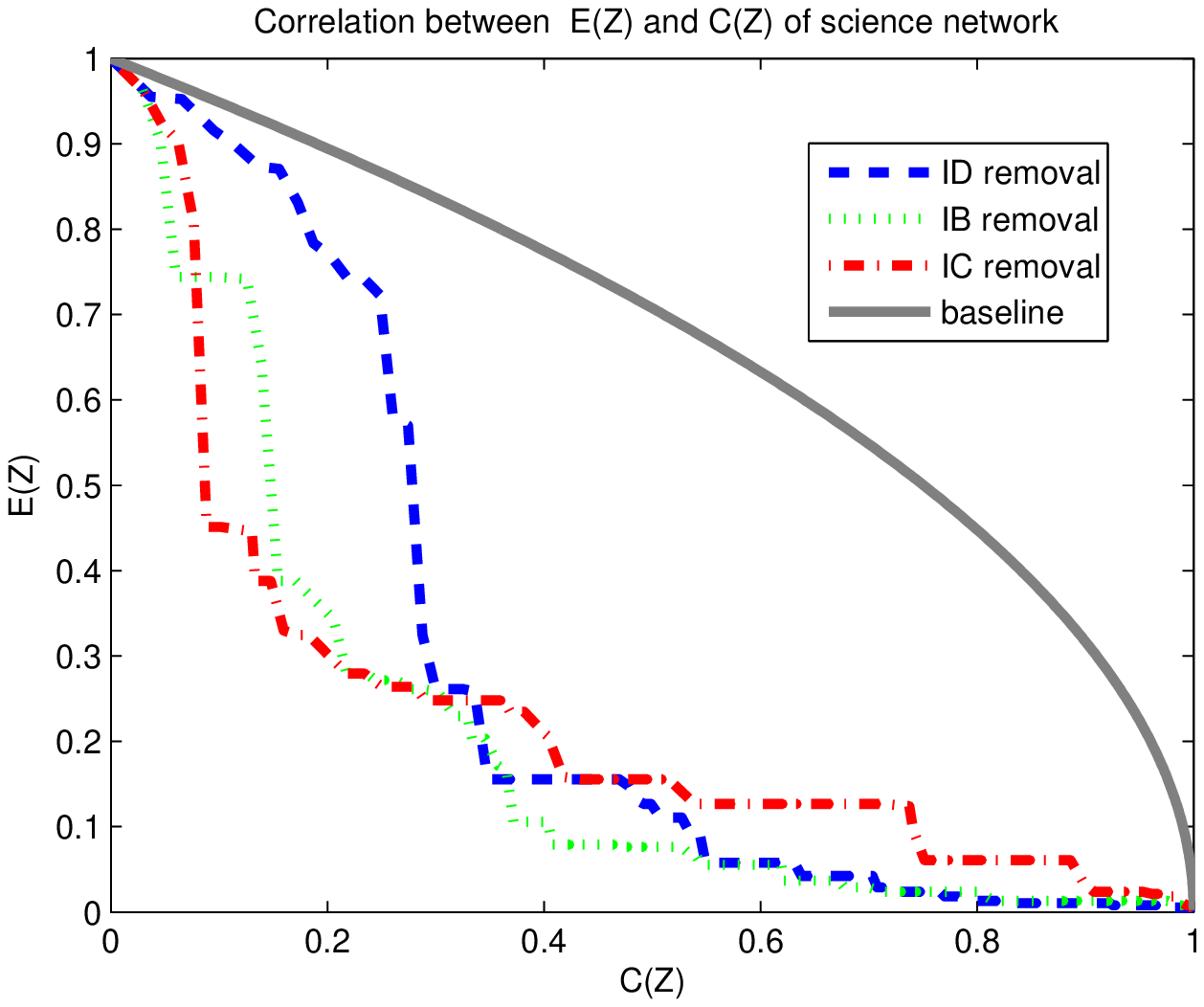}}
\caption{The Attack Effects with the Consideration of the Cost on the Selected Networks. Every point in the curves represents an attack. The X-axes are the responding attack cost $C(Z)$, the Y-axes are the responding attack effect $E(Z)$ for the same attacks. The baseline is the attack cost and the attack effect to the complete network. The points above the baseline mean better attacks which obtain better attack effect but with less cost percentage.}
\label{fig.RelationsPictures} 
\end{figure*}

Since we can not determine the robustness only by $C(Z)$ or $E(Z)$ respectively,
we plot the corresponding relationship between $C(Z)$ and $E(Z)$ in Fig. \ref{fig.RelationsPictures} for any given attack $Z$. To explore the relative robustness, we use the complete networks as the baseline. Any selective node attack would not make this kind of networks collapse unless all the nodes are removed. Therefore, the complete networks could be regarded as the most robust without the
consideration of the cost. But in the viewpoint of the cost, they may waste much resource on trivial nodes.

To explore the relationship between the robustness and the compactness or the average degree, we categorize the networks into two groups, which the first group has larger average
degree(that is, CSF-Compact, CSF-NonCompact and Polbooks) and the second with smaller average degree( i.e., Protein and Science).

As to the first group, Fig. \ref{fig.RelationsPictures} shows that all three networks have good resistance to ID attacks in contrast to the complete networks,
 because a majority of points are located above the baseline,
 meaning that ID attacks on these networks would pay larger percentage of cost but achieve less effects.
From the subfigure \ref{fig:subfig:a:r}, we can see that every curve of the CSF-Compact network does not differ much from the others and the majority parts of them locates above the baseline, meaning that this network can resist all three attack strategies, but for the subfigures \ref{fig:subfig:b:r} and \ref{fig:subfig:c:r}, the curves differ much, especially for \ref{fig:subfig:b:r}, most points of IB curve locate below the baseline, meaning that  the CSF-NonCompact network and the Polbooks network can not perform well for IB attacks, this phenomenon must be attributed to their smaller $\delta_b$ such that the important nodes can be removed with small cost. As to the second group, Fig. \ref{fig.RelationsPictures} shows that both networks do not perform well to all the attacks in contrast to the complete networks, since the curves are totally below the baseline. But owing to the compactness, the Protein network still performs better than the Science network.

Two check points are chosen to quantitatively demonstrate the attack effects on these networks.
The first check point is E(Z)=80{\%}. Tab. \ref{tab.80} lists the attack cost under ID, IB and IC for five networks.

\begin{table}\caption{C(Z) when E(Z)=80{\%}}
\begin{tabular}
{cccc}
\hline
Network &
ID&
IB&
IC\\
\hline
CSF-Compact&
{71.25{\%}}&
{70.77{\%}}&
{66.79{\%}} \\
{Polbook}&
{63.04{\%}}&
{48.30{\%}}&
{45.35{\%}} \\
{CSF-NonCompact}&
{68.78{\%}}&
{24.52{\%}}&
{53.53{\%}} \\
{Protein}&
{19.85{\%}}&
{19.77{\%}}&
{24.63{\%}} \\
{Science}&
{18.22{\%}}&
{5.51{\%}}&
{7.68{\%}} \\
\hline
\end{tabular}
\label{tab.80}
\end{table}

Tab. \ref{tab.80} shows that 20\% size of giant component of the CSF-Compact network would need to pay 71.25\% cost under the ID strategy, moreover, the gap between ID and IB is very small, the gap between ID and IC is also small. But for the CSF-NonCompact network, the gap between IB and ID achieves 44.26\%; as to the Polbooks network, the gap also achieves 14.74\%, meaning that both networks have community structures and can not resist all attack strategies. As to the Protein network and the Science network, because less average edges leads to fast collapses, less cost can achieve better attack effects.

The second check point is chosen at E(Z)=30{\%}. Tab. \ref{tab.30} lists C(Z) under ID, IB and IC strategies.

\begin{table}\caption{C(Z) When E(Z)=30{\%}}
\begin{tabular}
{cccc}
\hline
Network&
ID&
IB&
IC \\
\hline
{CSF-Compact}&
{92.15{\%}}&
{91.68{\%}}&
{92.84{\%}} \\
{Polbook}&
{78.41{\%}}&
{72.91{\%}}&
{75.58{\%}} \\
{CSF-NonCompact}&
{89.31{\%}}&
{35.54{\%}}&
{56.31{\%}} \\
{Protein}&
{54.38{\%}}&
{50.33{\%}}&
{54.30{\%}} \\
{Science}&
{29.24{\%}}&
{21.39{\%}}&
{20.19{\%}} \\
\hline
\end{tabular}
\label{tab.30}
\end{table}

Tab. \ref{tab.30} shows that the CSF-Compact network still is very robust, achieving 30\% size of giant component would spend over 90\% cost for all three strategies. As to the other networks, less cost means that great collapse has emerged.

The experimental results support the proposed ideas. We can further rationalize the ideas in a simple way. Here we assume that the size of giant component is still used to evaluate the attack effects and the cost is considered.

The process is demonstrated with a simple language as follows,

\textbf{Proposition 1: } The RD strategy is the worst attack strategy when the target networks do not collapse.

\textbf{Proof: } Assume that $Z|_{RD}$ is a set to represent an RD attack, and $E(Z|_{RD})=\|N\|-\|Z|_{RD}\|$, that is, the network does not collapse.  For any given $Z'$ and $\|Z'\|=\|Z|_{RD}\|$, because $E(Z') \leq \|N\|-\|Z'\| = E(Z|_{RD})$, and $C(Z') \leq C(Z|_{RD})$. i.e., the RD attack pays more cost but obtains worst effects. So the RD strategy is the worst attack strategy.

According to proposition 1, if we can construct a network that does not collapse under the RD strategy, then this network would be very robust under RD selective attack.

\textbf{Proposition 2: } Any connected network ($\|N\| \leq 3$) would have at least a collapse under all RD attacks except the complete networks.

\textbf{Proof: } Considering the connected network with $\|N\|=3$, when it is not complete, it would collapse under the RD attack. Since the connected network with $\|N\|=3$ is the most foundational case, it is easy to conclude that only the complete networks would never collapse under the RD attacks.

The complete networks would never collapse, but most networks is not complete. In the real world, it would be a massive resource waste to construct such complete networks. That is, the collapses are essential. A robust network for selective attack is not a network which never collapse, but a network resists the attacks with higher efficiency before the collapses and has endurance to collapse.

A selective attack strategy actually defines a sequence of attacks for a specific network. Each element in the sequence is a set, i.e., an attack.

\textbf{Proposition 3}: For a specific network, for any given measure if the attack strategy based upon this measure can generate an attack sequence which is the same as the sequence of RD, then this attack strategy is the worst before the collapse.

This proposition is obvious, it means that if a network is robust under the selective attacks, then this network would be: 1) the attack strategies based upon the importance measures are the same as RD. 2) The networks can resist more time, that is,the collapses would emerge later.

As to the scale-free networks, since ID is very similar to RD, we can use ID to approximate RD. Hence, proposition 3 can be rewritten as follows,

\textbf{Proposition 3'}: For a specific scale-free network, for any given measure, if the attack strategy based upon this measure can generate an attack sequence which is the same as the sequence of ID, then this attack strategy is bad before the collapse.

Some scale-free networks, such as the CSF-Compact network, have the compactness property, that is, if one node is important in a measure, often also important in another measure.

Considering the endurance before the collapses for the scale-free networks. If the exponent of degree distribution keeps constant, the relative important nodes, even the trivial nodes would gain more edges when the average degree increases. This fact means that the ID attack would be more difficult to divide the network.

According to the analysis above, it would be reasonable that the compact scale-free networks with higher average degree would resist the ID attack strategy, and simultaneously the other selective node attack strategies.

In general, this Letter advances the current ideas on the fragility of the scale-free networks and proposes that some scale-free networks could be robust under the selective node attacks when the networks are compact with a high average degree considering the cost of attacks. The experimental results on five networks validate this idea.

However, this Letter spawns more problems. First, what kinds of networks would be the most robust networks under a cost constraint? This problem is related to build robust social, biological and technological networks, or to destroy these networks, and probably be useful to understand the evolutionary principles of these networks. Second, what will happen when considering the vertices attacks? Third, whether there exists a phase transition for the average degree vs. the robustness of the networks? Fourth, which attack strategies would be the most efficient strategies for a specific network? etc. These problems will be investigated in the future works.

Besides, the conclusions in this Letter could be potential to explain the evolutionary principles of biological populations and social systems which obey the power law distributions. In these networks, the evolution of nodes can be owed to the new nodes which could be helpful of the stability of the whole networks, that is, the nodes would obey the law that ``the most stable nodes survive'', here the stability is the adaption. In the BA model, the nodes would obey the law that ``the strongest nodes survive'', here the power is the adaption. By comparing to these two models, we can conclude that the previous one provides one holistic viewpoint, and the BA model is based upon the individual viewpoint.


%
%

\acknowledgments
Thanks to Dr. J. L\"{u}, Dr. O. Burger and Dr. CL Zhou.
Supported by the National Basic Research Program of China (No.2007CB310804),
the National Natural Science Foundation of China (No.60803095),
the State key Laboratory of  Networking and Switching Technology(No. SKLNST-2010-1-04) and the SCUEC foundation(No. YZZ06025) .

\bibliographystyle{eplbib}

\begin{thebibliography}{10}
\expandafter\ifx\csname url\endcsname\relax\def\url#1{\texttt{#1}}\fi

\bibitem{a28}
\Name{Albert R., Jeong H. \and Barab\'{a}si A.-L.} \REVIEW{Nature
  }{406}{2000}{378}.

\bibitem{a18}
\Name{Holme P., Kim B.~J., Yoon C.~N. \and Han S.~K.} \REVIEW{Physical Review E
  }{65}{2002}{056109}.

\bibitem{a26}
\Name{Tan Y., Wu J., Deng H. \and Zhu D.} \REVIEW{Systems Engineering
  }{24}{2006}{1}.

\bibitem{a5}
\Name{Stanley W. \and Katherine F.} \Book{Social network analysis: methods and
  applications} (Cambridge University Press) 1994.

\bibitem{a6}
\Name{V\'{a}zquez A., Pastor-Satorras R. \and Vespignani A.} \REVIEW{Physical
  Review E }{65}{2002}{066130}.

\bibitem{1}
\Name{Barab\'{a}si A.~L. \and Albert R.} \REVIEW{Science }{286}{1999}{509}.

\bibitem{a9}
\Name{Olaf S.} \REVIEW{Complexity }{8}{2002}{56}.

\bibitem{a12}
\Name{Broder A., Kumar R., Maghoul F., Raghavan P., Rajagopalan S., Stata R.,
  Tomkins A. \and Wiener J.} \REVIEW{Computer Networks }{33}{2000}{309}.

\bibitem{a13}
\Name{Jeong H., Mason S.~P., Barab\'{a}si A.~L. \and Oltvai Z.~N.}
  \REVIEW{Nature }{411}{2001}{41}.

\bibitem{a14}
\Name{Jennifer A.~D., Richard J.~W. \and Neo D.~M.} \REVIEW{Ecology Letters
  }{5}{2002}{558}.

\bibitem{a15}
\Name{Newman M. E.~J., Forrest S. \and Balthrop J.} \REVIEW{Physical Review E
  }{66}{2002}{035101}.

\bibitem{a16}
\Name{Magoni D.} \REVIEW{Selected Areas in Communications, IEEE Journal on
  }{21}{2003}{949}.

\bibitem{a27}
\Name{Zheng B.} Tech. Rep. Tsinghua University PostDoc116-2008 (2009).

\bibitem{a29}
\Name{Zheng B., Wang J., Chen G., Jiang J. \and Shen X.} \REVIEW{Chinese
  Physics Letter }{28}{2011}{018901}.

\bibitem{a19}
\Name{Freeman L.~C.} \REVIEW{Social Networks }{1}{1978}{215}.

\bibitem{a21}
\Name{Brin S. \and Page L.} \REVIEW{Computer Networks and ISDN Systems
  }{30}{1998}{107}.

\bibitem{23}
\Name{Page L., Brin S., Motwani R. \and Winograd T.} Tech. Rep. Computer Science Department,
  Standford University (1998).

\bibitem{a23}
\Name{Jon M.~K.} \REVIEW{J. ACM }{46}{1999}{604}.

\bibitem{a2}
\Name{Newman M. E.~J.} \REVIEW{SIAM Review }{45}{2003}{167}.

\bibitem{a32}
\Name{Clauset A., Shalizi C. \and Newman M. E.~J.} \REVIEW{SIAM Review
  }{51}{2009}{661}.

\end{thebibliography}

\end{document}